\documentclass[12pt]{revtex4}
\usepackage{graphicx,color}
\usepackage{amsmath,amsfonts,amssymb,amsthm, bm}
\usepackage{subfigure}
\usepackage{pslatex}
\def\bea{\begin{eqnarray}}
\def\eea{\end{eqnarray}}
\def\be{\begin{equation}}
\def\ee{\end{equation}}
\def\nn{\nonumber}

\def\ph{\phi}

\def\th{\theta}

\def\p{\partial}

\def\p{\partial}
\def\be{\begin{equation}}
\def\ee{\end{equation}}
\def\tt{\tilde{t}}

\begin{document}

\title{Semiclassical corrections to the photon orbits of a non-rotating black hole}
\author{Swayamsiddha Maharana\footnote{current address: School of Mathematical and Physical Sciences, Macquarie University, NSW 2109, Australia.} ,  Arundhati Dasgupta ,\footnote{E-mail: arundhati.dasgupta@uleth.ca}}
\affiliation{Physics and Astronomy, Science and Academic Building \\ 4401 University Drive, \\University of Lethbridge,\\
Lethbridge T1K 3M4.\\}


\begin{abstract}
 In this brief article we discuss the corrections to the photon orbits of a non-rotating black hole due to semiclassical fluctuations of the metric. It is found that the photon orbit impact parameter differences with the critical impact parameter become of the order of the semiclassical fluctuations. We calculate the effect of the semi-classical fluctuations on the photon orbits and show that instead of circulating the black hole infinite number of times at the critical orbit, the photons bounce off the semiclassical geometry.
\end{abstract}
\maketitle

\section{Introduction}
The image of the black hole at the centre of the M87 Galaxy was obtained using VLBI and announced in a series of papers \cite{eht}. A very remarkable achievement based on data from 8 infrared interferometers, placed at various points on Earth, the image comprises of a central `shadow', surrounded by circular photon orbits. Whereas most galactic centre black holes have non-zero rotation parameters, the shape of the image differs from the non-rotating one only by 4\% \cite{eht}.  In this paper we begin the discussion of `quantum gravity' corrections for this image by studying the photon orbits which generate the non-rotating black hole image. We expect that our results can be easily extended to the rotating example.  Our calculations are valid for perturbations of the metric originating from any existing quantum gravity theories, however, we compute the numeric values of the corrections using the formulas found in \cite{adg}.
In \cite{adg}, semiclassical states in loop quantum gravity (LQG) had been used to study corrections to the classical metric. Whereas, the corrections are at the level of linear `perturbations' of the metric, the form of the corrections are non-polynomial in nature.  Whether the predictions in \cite{adg} are true or not have to be verified using experiments. We predict the corrections to the critical impact parameter, using explicit numerical values, these are very small $\sim 10^{-9}  M-10^{-66} M $ (M being the mass scale of the black hole) and probably can be verified in future images sensitive to distances between photon orbits and or interference fringes.  The lower limit of the range comes from the primordial blackholes whose radius is about $10^{3} $ Planck length and the upper limit of the range is for astrophysical solar mass black holes. 

In this article, we follow the calculations of \cite{chandra,luminet} for the metric of \cite{adg}  and show how the photon orbits will change due to the semi classical corrections. There is a critical impact parameter for photons, after which they are captured by the black hole's gravitational field. For photon geodesics with impact parameter greater than the critical value, the photon geodesics can escape back to the asymptotic, carrying information about the black hole with them. The photon geodesics can encircle the black hole n-times before escaping. The photon geodesic with the critical impact parameter encircles the black hole infinite times. As the number `n' increases, the difference of the impact parameter and the critical impact parameter decreases. What we find interesting about the classical results is that the difference of the photon orbit impact parameter with the critical impact parameter can be of the order of the semiclassical fluctuations of the metric for geodesics encircling the black hole only three times (n=3). This raises the question, how would the semiclassical fluctuations of the metric affect the classical analysis of these systems. We perform an explicit numerical calculation using the corrections of \cite{adg} to see the physics of the semiclassical fluctuations.  We also try to obtain an analytic expression for the photon orbit corrections. We find that for small black holes the effect on the orbits is rather drastic, but for astrophysical black holes the nature of the correction is to slightly change the absorption cross-section. For small black holes we identify an integer `n' which characterizes the maximum number of times the photons circle the black hole, as a quantum number fixed by the semiclassical scale of the system. This number is given as $2 n \pi \propto -\ln (\tt) $ , where $\tt$ is the semi-classical parameter, characterizing the scale of semi-classical fluctuations. The classical limit is when $n\rightarrow \infty$ and $\tt \rightarrow 0$. This n characterizes the maximum number of times the photon can rotate around a semiclassical black hole.  We report on the expected changes, but the photographic plate image construction is yet work in progress. {\it Note this calculation is highly restricted by the `semiclassical' linear perturbation techniques.} We expect that non-perturbative quantum gravity will show the correct equations for the nature of photon orbits around quantum black holes. For astrophysical black holes, the nature of corrections are very tiny, but yet can be detected in future experiments. 

In the following section, we discuss the nature of the geodesic corrections for generic perturbations of the metric. In the section following that, we compute exact numerical values of the corrections using the semiclassical metric of \cite{adg}. In the third section we conclude and discuss work for the future. 

\section{The corrections to the Geodesic and the Photon orbits}

We take the `semi classically corrected' Schwarzschild metric to be of the form
\bea
 ds^2  & = & -\left(1- \frac{r_g}{r} - \tilde{t} \ h_{tt} \right) dt ^2 + \tilde{t}\  h_{rt} \ dt dr  + \left\{\frac1{(1-r_g/r)} + \tilde{t} \ h_{rr} \right\} dr^2 + \left(r^2 + \tilde{t}  \ h_{\th \th} \right) d\th^2 \nn  \\
 && + \left(r^2 \sin^2\th + \tilde{t} \ h_{\phi \phi} \right) d\phi^2 
 \label{eqn:corrm}
\eea
$\tilde t$ is a semiclassical parameter, and $h_{ij}(t,r,\th,\phi)$ ($i,j=t, r, \theta,\phi$) are metric fluctuations which are coordinate dependent. These corrections can arise due to quantum gravity, quantum energy momentum tensor fluctuations of matter fields etc. $r_g= 2GM$ is the Schwarzschild radius.  The form of the corrections, and the fact that there is only one cross term $h_{tr}$ is motivated from the semi classical corrections obtained in \cite{adg}.

Using the calculations of \cite{luminet} and \cite{chandra}, we take the geodesics in the $\theta=\pi/2$ plane or the equatorial photon orbits and calculate their general behaviour. The geodesic equation up to $\mathcal{O}(\tilde{t})$ semi classical corrections is given as (where L is angular momentum, and E the energy of the system)
\begin{equation}
\label{eqn:geo_1}
    \left(\frac{dr}{ds}\right)^{2}+\frac{1}{g}\left(\frac{L^{2}}{\tilde{q}}+\frac{E^{2}}{f}\right)=0
\end{equation}
where the functions are appropriately defined as in \cite{adg}. 
\be
f = -\left(1- \frac{r_g}{r} - \tilde{t}\   h_{tt} \right)  \   \  \    \   \     g= \left\{\frac1{(1-r_g/r)} + \tilde{t} \  h_{rr} \right\}  \  \   \    \tilde{q}= \left(r^2 \sin^2\th + \tilde{t} \ h_{\phi \phi} \right) .\label{eqn:defn}
\ee  
Using the conservation of angular momentum equation i.e. $\tilde{q}\frac{d\phi}{ds}=L$ gives from equation (\ref{eqn:geo_1})
\begin{equation}
\label{eqn:geo_2}
    \frac{1}{\tilde{q}^2}\left(\frac{dr}{d\phi}\right)^{2}+\frac{1}{g}\left(\frac{1}{\tilde{q}}+\frac{1}{fb^2}\right)=0
\end{equation}

where `b' is the impact parameter defined as 
\be
b^2 = \frac{L^2}{E^2}.
\ee

We can then write the above equation using the explicit forms of the functions as defined in (\ref{eqn:defn}) 

\be
\frac{1}{(r^2 + \tilde{t} h_{\phi \phi})^2} \left(\frac{dr}{d\phi}\right)^2 + \frac{1-r_g/r}{1+ \tt h_{rr}(1-r_g/r)} \left(\frac{1}{r^2 + \tt h_{\phi \phi}} - \frac{1}{b^2(1- r_g/r - \tt h_{tt})}\right)=0.
\ee
Using $\tt$ as a small parameter, one can do binomial expansion of the following (Note as $r> 2M$ the factor $1-r_g/r$ is not zero, and can be used to factor out of the binomials and as the semiclassical parameter $10^{-9}>\tt >10^{-66}$ ).
 \bea
\frac{1}{r^4} \left(\frac{dr}{d\phi}\right)^2+ \left(1-\frac{r_g}{r}\right)\left(1+\tt \  h_{rr}\left(1-\frac{r_g}{r}\right)\right)^{-1}\left[\frac1{r^2}\left(1+\tt \ \frac{h_{\phi \phi}}{r^2}\right)^{-1} \right.&& \nn \\- \left. \frac1{b^2(1-r_g/r)}\left(1-\tt \ \frac{h_{tt}}{1-r_g/r}\right)^{-1}\right]\left(1+\tt \ \frac{h_{\phi \phi}}{r^2}\right)^{2}=0 .&&
\eea
Keeping order $\tt$ terms in the binomial expansions one gets:
\be
\frac{1}{r^4}\left(\frac{dr}{d\phi}\right)^2 + \left[\frac1{r^2}\left(1-\frac{r_g}{r}\right) -\frac{1}{b^2} \right]\left(1+ \tt \frac{h_{\phi \phi}}{r^2} - \tt h_{rr}\right)  - \frac{\tt}{b^2}\left(\frac{h_{\phi \phi}}{r^2}+ \frac{h_{tt}}{1-r_g/r}\right)=0 .\label{eqn:pot1}
\ee
In the above we have used Binomial expansion in powers of $\tt$. 
If we re-write the above in a convenient form we get
{\small
\be
\frac{1}{r^4}\left(\frac{dr}{d\phi}\right)^2+ \frac1{r^2}\left(1-\frac{r_g}{r}\right)\left(1+ \tt \frac{h_{\phi \phi}}{r^2} - \tt h_{rr}\right) = \frac{1}{b^2}\left(1+ 2\tt \frac{h_{\phi \phi}}{r^2} -\tt h_{rr} +\tt \frac{h_{tt}}{1-r_g/r}\right).
\ee} 

The above equation is of the form
\be
\frac{1}{r^4}\left(\frac{dr}{d\phi}\right)^2 = \frac{1}{b^2} H_1(r)- V_1(r)
\ee
where $V_1(r)=\frac1{r^2}\left(1-\frac{r_g}{r}\right)\left(1+ \tt \frac{h_{\phi \phi}}{r^2} - \tt h_{rr}\right)$ and $H_1(r)=\left(1+ 2\tt \frac{h_{\phi \phi}}{r^2} -\tt h_{rr} + \tt \frac{h_{tt}}{1-r_g/r}\right)$.
The equation will have a solution iff 
\be
\frac{1}{b^2} H_1(r)- V_1(r)\geq0.
\ee
From this we identify `the potential function' as
\be
\frac{1}{b^2}\geq \frac{V_1(r)}{H_1(r)}\geq V(r).
\label{eqn:pot}
\ee

where the potential function $V(r)$ to order $\tt$ is identified as 
\be
V(r)\equiv \frac{1}{r^{2}}\left(1-\frac{r_{g}}{r}\right)\left(1-\tilde{t}\frac{h_{\phi\phi}}{r^{2}}-\tilde{t}\frac{h_{tt}}{\left(1-\frac{r_{g}}{r}\right)}\right).
\ee

To find the extremum of the potential we take the derivative of the potential and set it to zero.
\be
\frac{\p V}{\p r}|_{r=r_c}=0.
\ee

That gives
\bea
\frac1r \left(-2 + \frac{3 r_g}{r}\right) \left( 1 - \tilde{t} \frac{h_{\phi \phi}}{r^2} - \tilde{t} \frac{h_{tt}}{1- r_g/r} \right)  \nn &&\\
+ \tilde{t} \left(1- \frac{r_g}{r}\right) \left[ 2 \frac{h_{\phi \phi}}{r^3} - \frac{\partial_r h_{\phi \phi}}{r^2} - \frac{\partial_r h_{tt}}{1-r_g/r} + \frac{h_{tt}}{(1-r_g/r)^2} \frac{r_g}{r^2}\right] && =0. 
\eea

At the zeroeth order the above gives the critical radius to be
$r_0= (3/2) r_g= 3M$. This is an `unstable' orbit, where the potential has a maximum. We next assume a correction to this critical radius which is given by
$$ r_c= r_0 + \tilde{t} \ \xi. $$
The correction can be solved as
\be
\xi= \frac1{9 r_g}  \left[ 2  h_{\phi \ph}(r_0) -r_0  \p_r h_{\phi \phi}  + r_0 r_g\frac{h_{tt}(r_0)}{(1-r_g/r_0)^2}  - r_0^3\frac{\p_r h_{tt}(r_0)}{1-r_g/r_0} \right].
\label{eqn:corr}
\ee

A limit on the value of the `impact parameter' can be found using the fact that $(dr/d\phi)^2 \geq 0$; as in (\ref{eqn:pot}) which is explicitly:

\begin{equation}
   \frac{1}{ b^{2}}\geq \frac{1-\frac{r_{g}}{r}}{r^{2}}\left(1-\tilde{t}\left(\frac{h_{\phi\phi}}{r^{2}}+\frac{h_{tt}}{1-\frac{r_{g}}{r}}\right)\right).
\end{equation}
Given that the potential has a maximum at $r_c$ this is same as
\be
b \leq (V(r_c))^{-1/2}.
\ee
Which shows that there is a  critical impact parameter $b_c = (V(r_c))^{-1/2}$.
This is the critical impact parameter, after which the photon is absorbed into the black hole, and cannot escape back to the asymptotics. As the potential as well as the critical radius is corrected, we get a new `inner disk' radius and hence a corrected absorption cross section for the black hole \cite{adg2}. 

As stated above, the photons reaching the black hole with an impact $\leq b_c$ are captured by the black hole. Thus there is a `hole' of radius $b_c$ a disc from within which light does not reach the observer. If we inspect the corrections, then they are tiny. From the discussions in \cite{adg}, one takes the semiclassical parameter, in a certain range, depending on the ratio of the length scale of the space-time and the Planck length. This range is 
$10^{-9} \leq\tilde{t} \leq 10^{-66}$, and therefore, the correction to the impact parameter, the disc radius and the absorption cross section of the black hole is very small. Given the resolution of the current image \cite{eht}, it shall be difficult to discern the semiclassical image corrections to the absorption cross section which is the area of the sphere with radius $b_c$. However, we make another observation based on \cite{luminet}  that the difference of impact parameters of photon orbits which encircle the black hole with the critical impact parameter $b_c$ is of the order of the semiclassical parameter, at n=3. This stems from the fact that classically \cite{luminet}
\be
b-b_c= 3.4823 \ M \exp(-\mu-2 n \pi)
\label{eqn:analytic}
\ee
where $n$ represents the number of times a photon encircles the black hole horizon. We observe that $\exp(- 2n \pi)\sim \tt$ for $n=3$ for $\tt \sim 10^{-9}$ a primordial black hole (\cite{adg}) and n=24 for $\tt\sim 10^{-66}$ which is a solar mass black hole. If this is true, then, the semiclassical fluctuations of the metric might be dominant at a much earlier stage, before the $b_c$ is reached. As $\tt$ sets the scale of the corrections to the above equation, we re-investigate the physics of the system. To investigate this, we solve for the geodesic equations of the semi-classically corrected metric.    
 We follow the methods of \cite{luminet,chandra} for the corrected metric (\ref{eqn:corrm}). We take the equations for (\ref{eqn:pot1}) and re-write in terms of $u=1/r$ and separate the classical and order $\tt$ parts of the equation.
We define the following quantities as
\begin{align}
G_{0}(u)       \equiv{} &u^{3}-\frac{u^{2}}{2M}+\frac{1}{2Mb^{2}} \\[2pt]
H(u)   \equiv{} &\frac{1}{2M}\left(u^{2}(1-2Mu)^{2}h_{rr}\left(\frac{1}{u}\right)-u^{4}(1-2Mu)h_{\phi\phi}\left(\frac{1}{u}\right)+\frac{h_{tt}\left(\frac{1}{u}\right)}{b^{2}(1-2Mu)}\right) \notag\\
                   &{}+\frac{1}{2M}\left(2\frac{h_{\phi\phi}\left(\frac{1}{u}\right)u^{2}}{b^2}-\frac{(1-2Mu)h_{rr}\left(\frac{1}{u}\right)}{b^{2}}\right) \\[2pt]
G(u)    \equiv{} &G_{0}(u)+\tilde{t}H(u) 
\label{eqn:define}
\end{align}
where $r=\frac{1}{u}$.\\
In the above, $H(u)= G_0(u) [(1- 2M u) h_{rr}  - u^2 h_{\phi \phi}] + 1/b^2 [h_{\phi \phi} u^2+ h_{tt}/(1-2Mu)]$. 
Substituting $r=\frac{1}{u}$ in the geodesic equation (\ref{eqn:pot1}) gives
\begin{equation}
\label{eqn:diff_1}
    \left(\frac{du}{d\phi}\right)^{2}=2MG(u)
\end{equation}
$G_0(u)$ is a cubic, and has three roots, in the classical limit; $u^0_1, u^0_2, u^0_3$. For which $u^0_1<u^0_2 <u_3^0$ and $u^0_1 <0$. From \cite{luminet} these are taken as 
\be
u_1^0= -\frac{Q_0-P_0 + 2M}{4 M P}  \  \   \    \    u_2^0 = \frac{1}{P_0}  \   \    \    \  u_3^0 = \frac{Q_0+P_0-2M}{4 M P_0} 
\ee

where $Q_0^2= (P_0-2M)(P_0+6M)$ ; $P_0$ being the location of the Periastron; and the  impact parameter is solved as by setting $G_0(u)=0$
 \be
 b_0= \frac{P_0^3}{P_0-2M}.
 \ee
 For the impact parameter at $b=b^0_c$, $P_0= Q_0=3M$ (unless stated otherwise $x^0$ or $x_0$;  physical quantities $x$ labelled with 0; represents a classical number) . 
  In the $G_0(u)$ function, without the semiclassical fluctuations, at $P_0=1/3M$ there is a double root. When we try to solve Equation (\ref{eqn:diff_1}) the integral of $1/\sqrt{G_0(u)}$ in the $\tilde{t}=0$  limit (or classical limit) can be approximated as $\sim \int du/(u-u_2)$, which obviously has a logarithmic divergence at $u=u^0_2=1/3M$. However, with the introduction of a shift from this as $u_2^0=1/[3M(1+\delta)]$ ($\delta$ being a small number) the degeneracy of the roots is broken and the integral is no longer divergent at $u=u_2$, but as expected the infinity is regulated as $\ln \delta$ which diverges as $\delta\rightarrow 0$.     
Due to the corrected form for $G(u)$, we take the order $\tilde t$ corrections to the above roots. 
Using the same derivation as in \cite{luminet,chandra} we take the Periastron distance $P$ as a function of the second root $u_2=1/P$. The periastron distance in the corrected and uncorrected geodesic are given as $P$ and $P_{0}$ respectively. Only $\mathcal{O}(\tilde{t})$ corrections are considered to the periastron distance. So, the correction is to linear order in $\tt$ 
\begin{equation}
    u_{2}\equiv \frac{1}{P}=\frac{1}{P_{0}}+\tilde{t}\nu
\end{equation}
and $u_{2}^{0}\equiv\frac{1}{P_{0}}$.  As $P_0=r_0$ the correction to $u_2$ is taken as $\nu= -\xi/r^2_0=-\xi/(3M)^2$, where $\xi$ is defined in Equation(\ref{eqn:corr}). If we deviate away from the $3M$ then $\nu$ has a correction proportional to the deviation $\delta$ too.
Since $P$ is the periastron distance (the closest point or the turning point of the trajectory as the particle scatters off the black hole)
\begin{equation}
\label{eqn:geo_3}
    \frac{1}{2M}\left(\frac{du}{d\phi}\biggr\rvert_{u(\phi)=u_{2}}\right)^{2}=0=G(u_{2}=u_{2}^{0}+\tilde{t}v).\\
    \end{equation}
    We can solve for the $\nu$ by observing that the corrected $G(u)$ can be written as
\be
G(u_2^0 +\tilde{t} \nu)= G_0(u_2^0) +\tilde{t} \nu G'(u_2^0) + \tilde{t} H(u)=0.
\ee
If in addition we assume that the impact parameter is corrected upto $\tilde{t}$ as $b=b_0 + \tilde{t} \tilde{b}$, one gets:
\be
(u_2^0)^3 - \frac{(u_2^0)^2}{2M} + \frac{1}{2M b_0^2} - \frac{\tilde{t} \tilde{b}}{M b_0^3} + \tilde{t} \nu G'(u_2^0) + \tilde {t} H(u_2^0)=0.
\ee
When $u_2^0=1/3M$ the system has a double root, and $G'(u_2^0)$ is zero. The above equation can be used to solve for 
$\tilde{b}$ as ($u_2^0=1/(3M)$),
\be
\tilde{b}= M b_0^3 ~H(u_2^0).
\ee
When $u_2^0\neq 1/(3M)$ one gets
\be
\tilde{b} =M b_0^3 (G'(u_2^0)~\nu+ H(u_2^0)).
\ee
 which also can be written as
 \be
 \frac{\tilde b}{M b_0^3} - \nu G'(u_2^0)= H(u_2^0)
 \label{eqn:b}
 \ee
It is difficult to compute the integral of the differential equation (\ref{eqn:diff_1}) analytically as the semiclassical function has a quintic. Numerical values of the integral for the semiclassical corrected equation differ from the classical integral at the order of $\tilde{t}$ which is going to interfere with the classical calculation of the impact parameter. We have used explicit expressions for the corrections to the metric from \cite{adg2}. The details of the expression can be found in the next section. Here we present the numerical calculations to motivate the analytic calculations.  
We define
\be
\phi_{\infty}=\frac1{\sqrt{2M}} \  \int_{0}^{u_2} \frac{du}{\sqrt{G(u)}}
\ee
where the photon traverses from $r=\infty, u=0$ to $r=3M(1+\delta) + \tt \nu; u=u_2$ the periastron, and the angle $\phi$ changes from $0$ to $\phi_{\infty}$. In the table, we have the description of columns (i)  the value of the semiclassical parameter $\tt$. (ii) The deviation from the maximum $3M$, the $\delta$ (iii) The exact numerical value of the integral or $\phi_{\infty}$ obtained using Mathematica (iv) The value of the integral without the order $\tt$ terms in G(u) (v) Comparison of the classical value with the semiclassical one labelled as $\Delta\phi_{\infty}$
  \begin{center}
\begin{tabular}{ |c|c|c|c|c|} 
\hline
$\tilde{t}$ & $\delta$ & Exact numerical integral $\phi_{\infty}$& Integral without semiclassical correction $\phi^0_{\infty}$ &$\Delta \phi_{\infty}$ \\
\hline
$10^{-9}$ & $10^{-8}$ & $19.588629507213$ & $19.588629510149$ & 2.936 $\times 10^{-9}$\\ 
$10^{-20}$ & $10^{-18}$ & $42.61448042675600592$ & $42.61448042675600334$ &$\sim 10^{-15}$\\ 
$10^{-12}$& $10^{-11}$ & $26.496384771569947$ & $26.496384775811023$& $\sim 10^{-9}$\\ 
$10^{-9}$& $10^{-7}$ & $17.28604453548423$ & $17.2860445371548$&$\sim 10^{-8}$ \\
$10^{-12}$& $10^{-10}$ & $24.193799682753037$ & $24.193799682936977$& $\sim 10^{-10}$\\
$10^{-15}$& $10^{-12}$ & $28.7989698687928421$ & $28.7989698687930655$ &$\sim 10^{-13}$\\
\hline
\end{tabular}
\end{center}
The exact integral differs from the one without the semiclassical parameter almost to order $\tilde{t}$. How would these changes affect the image of the event horizon?  For the purposes of this paper we see how the impact parameter as a function of $\delta$ and therefore the scattering angle  is modified due to the semiclassical corrections in a analytic formula.
As the classical formula is obtained analytically (Equation(\ref{eqn:analytic})), and we try to obtain a similar analytic formula  for the semiclassical case too in the following.
In Equation(\ref{eqn:analytic}), we can see what our numerical table suggests. Let us say in Equation(\ref{eqn:analytic}), $n=3$, then $b_3-b_c\approx 2.39 \times 10^{-9} M$ for $\mu=1 \ {\rm radians}$ and therefore for a photon traversing back to the photographic plate after encircling the black hole three times, the semiclassical fluctuations will be relevant for the above formula. How does the corrections to $\phi_{\infty}$ observed in the table above due to inherent semiclassical fluctuations of the metric affect the physics of the system? For that, we have to solve the equation analytically and obtain a functional relation between $\phi_{\infty}$ and the impact parameter $b$. However given the quintic nature of the function $G(u)$ analytical computations could not be obtained, neither did MAPLE or MATHEMATICA give us analytic results. We therefore obtained an approximate value for the integral analytically which we discuss next. 

To estimate analytically what the new physics might be, we approximate the square root using a linear order in $\tt$ expansion.
From equation(\ref{eqn:diff_1})
if we observe the structure of the $G(u)$, then it is of the form $H(u)= G_0(u) [(1- 2M u) h_{rr}  - u^2 h_{\phi \phi}] + 1/b^2 [h_{\phi \phi} u^2+ h_{tt}/(1-2Mu)]$. The first term has a double root at $1/3M$, however the second term has one root. 

To see the nature of the correction at order $\tilde{t}$ we make a analytic calculation based on the following discussions and approximations: 

At $u=u_2=u_2^0 + \tilde{t}\nu$ we find that we can write the $G(u)$ as $G_0(u_2) + \tilde {t} H(u_2) = G_0 (u_2^0) + \tilde t \nu G'(u_2^0)  - \tilde{t} \frac{\tilde{b}}{M b_0^3}+ \tilde t H(u_2^0)$, the first term by itself is zero, and the combination of the $\tt$ terms cancel each other. Thus we can use this split of the terms to obtain the function around $u\approx u_2$, and in general. (Note that this discussion is true if and only if $u_2 \neq 1/3M$.) This allows an approximation to the integral as a `binomial expansion' in the small parameter $\tilde {t}$. We use the notation $ H(u)+ \nu G'(u_2) - \frac{\tilde{b}}{M b_0^3}= \tilde{H}(u)$. 
\bea
\phi_{\infty}&= &\frac{1}{\sqrt{2M}} \int_{0}^{u_2} du \frac{1}{\sqrt{{G_0}(u)}} \left(1-\frac{\tilde{t}}{2}\frac{{\tilde{H}(u)}}{{G_0}(u)}\right)\\
&=& 2\left(\frac{P_0}{Q_0}\right)^{\frac{1}{2}}(K(k_0)-F(\zeta_{\infty},k_0))+\frac{1}{\sqrt{2M}}\int_{u_2^0}^{u_2} du \frac{1}{\sqrt{{G_0}(u)}} -\frac{\tilde{t}}{2\sqrt{2M}}\int_{0}^{u^0_{2}}du\,\frac{\tilde{H}(u)}{({G_0}(u))^{\frac{3}{2}}} \nn \\
\label{eqn:phi_1}
\eea
where 
\begin{eqnarray}
Q_0^{2}&\equiv&(P_0-2M)(P_0+6M)\\
k_0^{2}&\equiv&\frac{Q_0-P_0+6M}{2Q_0}\\
\sin^{2}\zeta{\infty}&\equiv&\frac{Q_0-P_0+2M}{Q_0-P_0+6M}
\end{eqnarray}
we put the integral in the third term as $I(u_2)$ which gives additional contribution to the solution for $\phi_{\infty}$, apart from the contribution from the Elliptic terms. The second term almost remains constant over the interval and we label that as $E_{1}$.

In the event that $\delta \approx t^{1/2}$ or smaller the $I(u_2)$ term contributes non-trivially and we investigate this in a separate subsection. Setting $\phi_{\infty}= \pi/2 + \mu/2$ where $\mu/2$ is the angle of scattering, one gets an equation for the impact parameter $b$. We try to solve the integral in the regime that the periastron is very close to $3M$. 
 
The requirement that $P_0>3M$ comes from the restriction $u_1<u_2<u_3$ which gives a condition $Q_0+ P_0 -6M>0$ \cite{chandra}. This is valid in the semiclassical calculation too.

Let 
\begin{equation}
\label{eqn:3M_1}
P_0=3M(1+ \delta)
\end{equation}

where as previously $ \delta$ is a small number above the mass of the blackhole. The equation which relates $\delta$ to $\phi_{\infty}$ gets corrected in the semiclassical approximation.
Hence, substituting equation (\ref{eqn:3M_1})  in equation (\ref{eqn:phi_1}) and then exponentiating both sides one gets 
\begin{equation}
\label{eqn:correction}
3.21\  \exp\left(-\phi_{\infty}\right)= 3.21 \exp\left(-\frac{\pi}{2}-\frac{\mu}{2}\right)=\delta \ \exp\left(\frac1{\sqrt{2M}} \left(-E_1 + \tt \frac{I(u_2)}{2}\right)\right).
\end{equation}
 
The interesting aspect of this calculation is that  for any $\mu$ as the number of cycles increases, very soon the order of the corrections become comparable with the semiclassical corrections to the Periastron.  Notice in the above that the integrals $E_1$ and $I(u_2)$ depend on the $\delta$ and this is a transcendental equation and cannot be solved if the exact form of the semiclassical corrections to the metric is not known. In the next section we use the semiclassical metric of \cite{adg2} to calculate the exact form of the above equation. 

As in the classical case, the impact parameter is defined as follows:

\begin{equation}
\frac{1}{b^{2}}=\frac{1}{27M^{2}}\left(1- 3\delta^{2}\right)\left(1-\tt u_2^{0 \ 2} h_{\phi \phi}(u_2^0) -\tt \frac{h_{tt}(u_2^0)}{1-r_g u_2^0}\right) = \frac{1}{27M^{2}}\left(1- 3\delta^{2}\right)\left(1-\tt a(u_2^0)\right)\end{equation}

where we have introduced $\tilde a(u_2)=u_2^{0 \ 2} h_{\phi \phi}(u_2^0) +\frac{h_{tt}(u_2^0)}{1-r_g u_2^0} $ to make the calculation easier. Thus
 
\be
b= 3 \sqrt{3} M \left(1+ \frac32 \delta^2\right)\left(1+\frac{\tt}{2} a(u_2^0)\right).
\label{eqn:corrf}
\ee
Next if we ignore the contributions from $E_1$ and $I(u_2)$. 
\be
b= 3 \sqrt{3} M \left( 1 + 15.48 \exp\left(-\pi -\mu \right) \right)\left(1+\frac{\tt}{2} a(u_2^0)\right).
\label{eqn:corr23}
\ee
for $\delta\sim \tilde{t}^{1/2}$, we find that the orbits differ from the critical one by an amount equal to the quantum fluctuation, as that being $\tilde{t}$ dominates over $ \delta$. 
If the images are eventually sensitive to be able to differentiate the $b_n$ th orbit from the $b_{n+1}$ th orbit, then the presence of quantum fluctuations will be detectable. In the next section we calculate the $a(u_2), I(u_2), E_1$ using the metric of \cite{adg}.
We can write Equation (\ref{eqn:corr23}) as 
\be
b-b^0_c = 3.48 ~M \exp(-\mu) \left(1-  \frac{\tt}{2} a(u_2^0)\right).
\ee

\begin{figure}
 \includegraphics[scale=0.3]{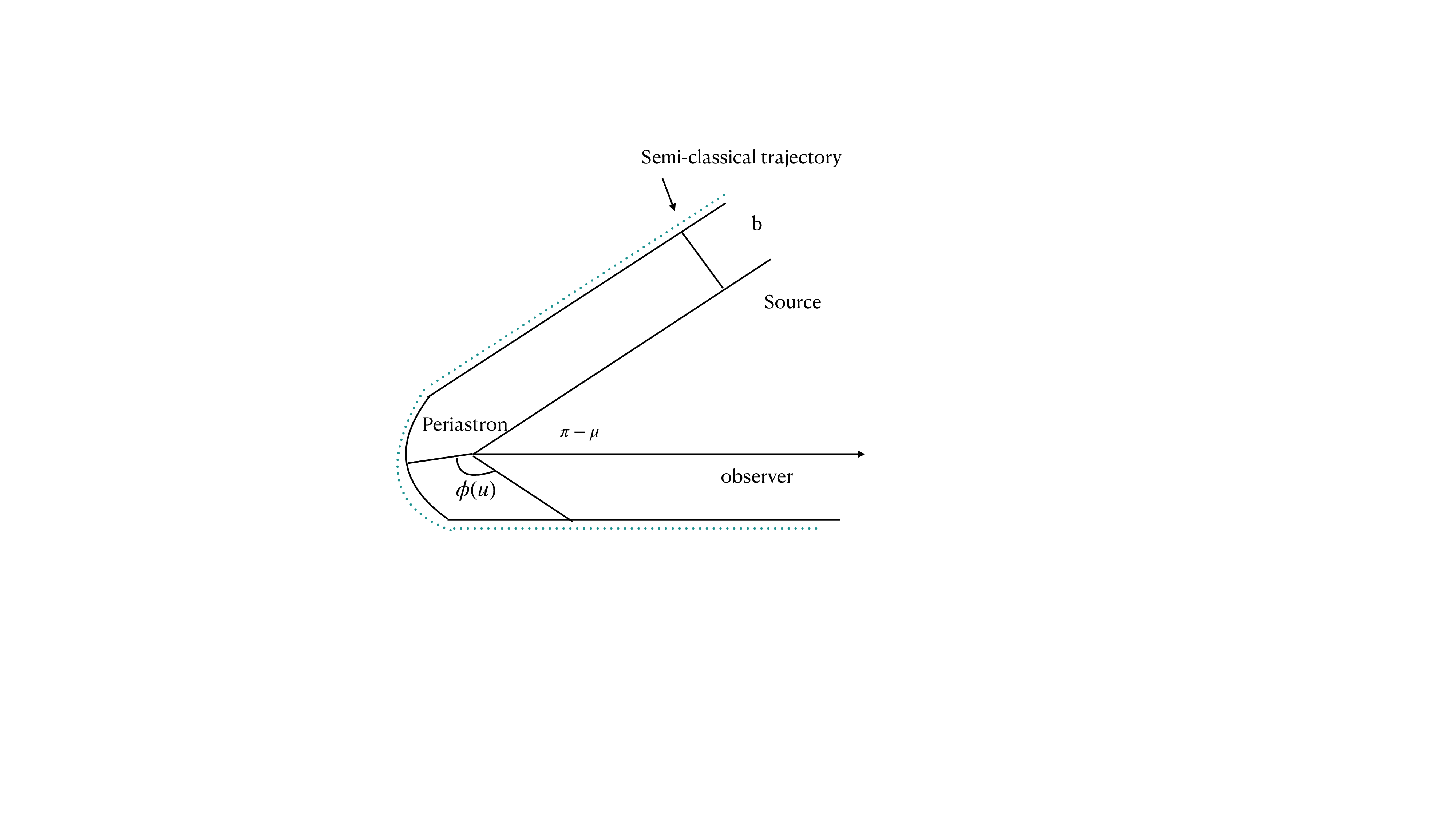}
 \caption{ The impact parameter increases due to semiclassical physics, but $\phi_{\infty}$ changes in a complicated way as a function of $\delta$}
 \end{figure}

As is obvious from the above, for $\tt \approx 10^{-9}$ a semiclassical fluctuation will have the same impact parameter as the one encircling the black hole thrice and reaching asymptotic region. However, $\tt \approx 10^{-9}$ only for primordial black holes with horizon of the order of $10^3$ Planck lengths. For solar mass black holes this $\tt\approx 10^{-66}$ here the semi-classically corrected impact parameter would be of the order of the impact parameter of the 24th orbit.

To find an explicit numerical values of the corrections, we use the form of the semiclassical metric as obtained in \cite{adg}. The functional form of the solution of the equation (\ref{eqn:correction}) can only be obtained after we have found the integrals $E_1$ and $I(u_2)$.

\subsection{Nature of the potential obtained from LQG Coherent States}

The semiclassical corrections to the non-rotating black hole were predicted in \cite{adg1}, computed in \cite{adg} and then discussed in details regarding their usefulness in making observations in a gauge invariant context in \cite{adg2}. Here we briefly give explicit expressions for the corrections to the metric as predicted in \cite{adg}. The Schwarzschild metric is well known in spherical coordinates, but the time slicing is not very convenient for the LQG graph embedding. If one takes the Lemaitre coordinates as in \cite{adg1}, then the time slices are flat. The slices also extend into the horizon up to the singularity such that one can address questions about singularity resolution as in \cite{adg3}.  The coherent states are obtained in this slicing, and semiclassical corrections to the metric are computed in these using techniques of LQG. The corrections to the flat metric in the three slices are as shown in \cite{adg}
\be
q_{\rm corr}^{ab}= q^{ab}\left(1+ 2 \ \tilde t \ f\left(\frac{P_{e_r}}{S_{e_a}}\right) \right)
\ee
where 
\be
f(x)= \frac1x\left(\frac1x-\coth(x)\right)
\ee

and $q^{rr}=1,  q^{\theta \theta}= 1/r^2,  q^{\phi \phi}= 1/(r^2\sin^2\theta)$ and $P_{e_a}$ is the appropriate LQG momenta for edges $e_a$ of a three dimensional graph, embedded in the time slices. In LQG the phase space is defined on a discretization of the three manifold in which the canonical degrees of freedom are defined. The three metric $q_{ab}$ is re-written in soldering forms $e_a^I$ (a=1,2,3 is the index on the three manifold world volume and I=1,2,3 is the tangent space index), such that $e_a^I e_b^I=q_{ab}$ a set of densitized triads $E^a_I= \sqrt{q} e^a_I$ when smeared over two dimensions surfaces which comprise the discretization of the three manifold are the above momenta. In the following the $e_a$ is the labelling of the edges in the `a' direction in three dimensions. The densitized triads are smeared over two dimensional surfaces $S_{e_a}$ (the details of this can be found in \cite{adg1,adg3}). These surfaces form the `dual' to the graph embedded in the three spatial slicing of the Lemaitre metric. 
\be
P^I_{e_{a}}= \int_{S_{e_a}}   * E^{I } 
\ee

and one can use a `gauge invariant' combination (inner product in the tangent space index $I$)
\be
P_{e_a}= \sqrt{ P^I_{e_a} P^{I}_{e_a}}.
\ee
For the `momenta' induced on the graphs, one obtains ($S_{e_a} \rightarrow 0$):

  \bea
  \frac{P_{e_r}}{S_{e_r}} & = & \frac{r^2 \sin\theta}{r_g^2}\\
  \frac{P_{e_\theta}}{S_{e_{\theta}}} & = & \frac{r\sin\theta}{r_g} \\
  \frac{P_{e_\phi} }{S_{e_{\phi}} }&=& \frac{r}{r_g} .
  \eea

If one sees what these are, they are the gauge invariant densitized triads of the LQG multiplied by $1/r_g$ the Schwarzschild radius to make the quantity dimensionless.

From the Lemaitre coordinates we make a transformation back to the Schwarzschild coordinates to obtain a `semi-classically corrected' metric.  Details of this can be found in \cite{adg},

The transformed semi classically corrected perturbations are:
\bea
h_{tt} & = &   - 2  \frac{r_g}{r} f\left(\frac{P_{e_r}}{S_{e_r}}\right)  = -2  r_g^3 u^3\left(r_g^2 u^2 - \coth\left(\frac{1}{r_g^2 u^2}\right)\right)\nonumber\\ 
  h_{rr} & = & -2 \frac{1}{(1-r_g/r)^2} f\left(\frac{P_{e_r}}{S_{e_r}}\right)= - 2 \frac{r_g^2 u^2}{(1-r_g u)^2} \left(r_g^2 u^2 -\coth\left(\frac1{r_g^2 u^2}\right)\right)  \nonumber \\
h_{\th \th} &= &  -2  r^2     f\left(\frac{P_{e_{\th}} }{S_{e_{\th}}}\right) = -\frac{2 r_g}{u}  \left(r_g u - \coth\left(\frac{1}{r_g u}\right)\right)     \nonumber \\    
h_{\phi \phi} &= & -2 r^2 \sin^2 \theta ~  f\left(\frac{P_{e_{\phi}} }{S_{e_{\phi}} }\right)= -2 \frac{r_g}{u}  \left( r_g u - \coth\left(\frac{1}{r_g u}\right)\right)
\label{eqn:exp}
\eea 

where we have taken $1/r=u$ and set $\sin\theta=1$ in the formulas.

Hence, the critical radius is given as (\ref{eqn:corr}), we find the explicit values of the functions at $r_0= 3M$,  as

\be
h_{tt}= 0.343  \   \   \   \     \partial_r h_{tt}= \frac{-.20734}{M}  \   \   \    \   \    h_{\phi \phi}= 5.2574 M^2  \   \    \    \    \     \    \    \partial_r h_{\phi \phi}= 3.096 M
\ee

  and 
\begin{equation}
\xi= 2.030 \ M.
\end{equation}

Given the critical radius, one can compute the critical impact parameter, beyond which the light rays get absorbed by the black hole. The light rays which fall on the black hole at this radius, follow an unstable radian geodesic. 
Light rays which are incident on the black hole with impact parameter bigger than the critical impact parameter escape back to the asymptotic, but they can encircle the black hole n-number of times, before emerging as discussed in the previous section. 
Here the critical impact parameter is given as
\begin{equation}
b_c= 3\sqrt{3} M \left( 1 + 0.8066  \  \tilde{t}\right).
\end{equation}

For the semiclassical corrections mentioned in this section we finally compute the impact parameter for the $n$-th orbit as : 
Given (\ref{eqn:corr});

\begin{equation}
\tilde a\left(\frac{1}{3M(1+\delta)}\right)=-.2389+ 0.2004 (3 \delta)
\end{equation}
Using equation (\ref{eqn:corrf}) 
\begin{equation}
b= 3 \sqrt{3}\  M + \frac{\sqrt{3}}{2} (9M \delta^2) + 4.1895  \ \tilde {t} \ M - 3.509 \ \tilde t \ 3M\delta
\end{equation}
where we have ignored the integrals $E_1$ and $I(u_2)$.  Note that the order $\tt$ term is the contribution to the impact parameter.

This discussion is relevant for $\delta > \tilde{t}$, but as we found before, about the 24th orbit and higher for $\tt \sim 10^{-66}$ , the difference of the impact parameter and the critical radius is of the order of the semiclassical parameter for solar mass black holes. Thus the question is what happens 25th orbit onwards? Next we try to compute the integral in Equation (\ref{eqn:integral}), with an explicit formula to see the corrections.  Note that in the Equation for $\phi_{\infty}$ as the zeroeth order term is $\ln(\delta)$ at order $\tt$ we keep only the terms which can compete with the same i.e. $\tt /\delta; \tt \ln\delta $ and ignore terms which contribute as $\tt \delta$ and higher.  We compute: 

\be
I(u_2) = \int_{0}^{u^0_2}d u \,\frac{\tilde{H}(u)}{({G}_0(u))^{\frac{3}{2}}}.
\ee

This using the definitions in Equations (\ref{eqn:define}, \ref{eqn:phi_1}) we find the above as
\bea
\int_0^{u^0_2 }\frac{d u}{G_0(u) ^{3/2}} \Bigg[ \Bigg.
\frac{1}{2M}\left(u^{2}(1-2Mu)^{2}h_{rr}\left(\frac{1}{u}\right)-u^{4}(1-2Mu)h_{\phi\phi}\left(\frac{1}{u}\right)+\frac{h_{tt}\left(\frac{1}{u}\right)}{b^{2}(1-2Mu)}\right) && \nn \\
                   {}+\frac{1}{2M}\left(2\frac{h_{\phi\phi}\left(\frac{1}{u}\right)u^{2}}{b^2}-\frac{(1-2Mu)h_{rr}\left(\frac{1}{u}\right)}{b^{2}}\right) \Bigg. \Bigg] - \int_{0}^{u_2^0} \ du \ \frac{\tilde{b}/M b_0^3-\nu G'_0(u_2^0)}{(G_0(u))^{3/2}}&&
 \label{eqn:tilda}
\eea

We concentrate on the first integral; by collecting the terms proportional to $h_{\phi \phi}$ and $h_{rr}$ one can factorize one power of $G(u)$. One gets the integrals as
\bea
\frac{1}{2M} \int_0^{u^0_2 }\Bigg[\Bigg.\frac{(1-2Mu)h_{rr} \left(\frac{1}{u}\right)}{\sqrt{G_0(u)}} - \frac{u^2 h_{\phi \phi}}{\sqrt{G_0(u)}} + \frac{u^2}{b^2} \frac{h_{\phi \phi}}{(G_0(u))^{3/2}} &&\nn\\
+ \frac{1}{b^2} \frac{h_{tt}}{(1-2M u) (G_0(u))^{3/2}} \Bigg. \Bigg] ~d u &&
\label{eqn:integral1}
\eea
The first two terms in the integral are not divergent as functions of $\delta$ where $P=3M(1+\delta)$ , but the second two terms have singular behaviour with $\delta$. In our analysis we keep terms which are divergent as functions of $\delta$ to solve the equation analytically. We compute them explicitly in the next section, but the divergence comes from the $\delta$ dependence in the Elliptic integral, as well as the dependence on $u-u_2\sim e$ from the power of $(G_0(u))^{3/2}$. However as we show in the appendix this potential divergence is cancelled in the total $I(u_2)$ due to contributions from the second term of Equation(\ref{eqn:tilda}).

From Equation(\ref{eqn:integral1}) the two terms which donot cancel are ($2M $ factored out) 
\be
\int_0^{u^0_2 } \frac{1}{b^2} \frac{ \left(u^2 h_{\phi \phi}  + h_{tt}/(1-r_g u)\right) d u}{(G_0(u))^{3/2}} .
\ee
Using the explicit formulas for the $h_{\phi \phi}$ and $h_{tt}$ from the equations (\ref{eqn:exp}), one gets the integral as
\be
\frac{2}{b^2} \int_0^{u^0_2} \frac{\left(r_g u - 2 r_g^2 u^2 + 2 r_g^3 u^3 - r_g^5 u^5\right) du}{(1-r_g u)(u^0_1-u)^{3/2} (u^0_2-u)^{3/2}(u-u^0_3)^{3/2}} .
\label{eqn:integral123}
\ee
We have approximated the $\coth(1/r_gu)$ and $\coth(1/r_g^2 u^2)$ as 1 in the above as the value of the function varies from $1.00$ to $1.023$ in the domain of definition of $u$. The corrections will be proportional to $\exp(-1/r_g u)$ and $\exp(-1/r_g^2 u^2)$ in the integrand and can be ignored at this level of the approximation. The explicit form of the integral is given in the Appendix, with even the zeroeth form of the roots $u^0_1$
$u^0_2 $ and $u^0_3  $  obtained upto quadratic powers of $\delta$ when $P_0=3 M(1+ \delta)$ plugged into the formula. Here $\delta$ is a dimensionless number to facilitate the calculation. From the terms in the appendix, we find that the terms proportional to 1/e, where $x=e$ is set as the lower limit of the integral. as well as inverse powers of $\delta$ dominate.
We find the equation to be from the Appendix Equations (\ref{eqn:I(u)}) for $I(u_2)$ and (\ref{eqn:E_1}) for $E_1$ in Equation (\ref{eqn:correction})
\bea
3.21 \exp(-\phi_\infty)& = & 3.21 \exp(-\pi/2 -\mu/2) \\
&= &\delta^{1+0.0203 \tt } \exp\left(+ \frac{0.47 \tt^{1/2}}{(0.67 \delta +0.225 \tt)^{1/2}} +0.23 \tt +1.712 \frac{\tt}{\delta}\right)\\
&=& w(\delta)
\eea

\begin{figure}
 \includegraphics[scale=0.5]{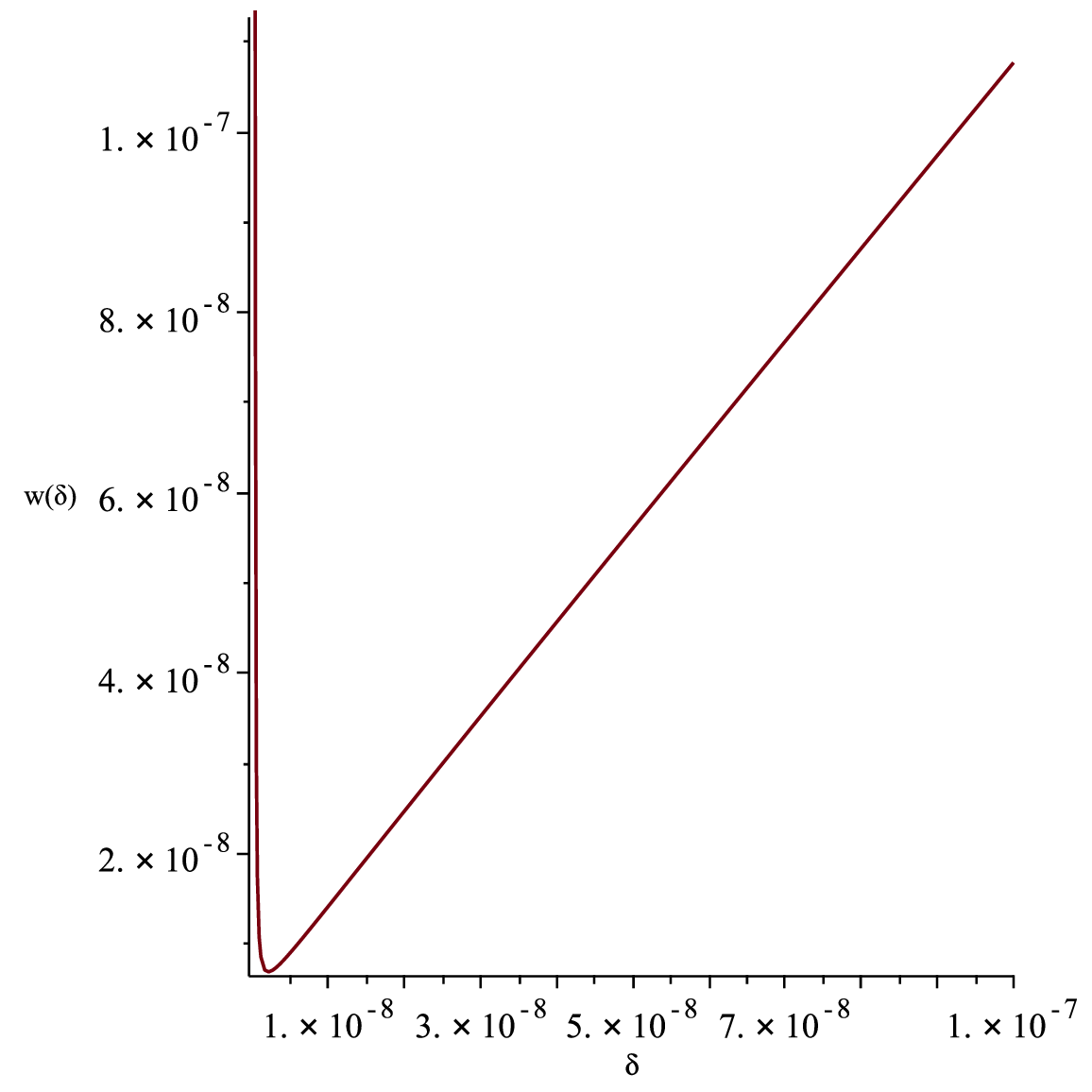}
 \caption{ $\tt \sim 10^{-9}$}
 \label{fig:graph1}
 \end{figure}

For $10^{-9}=\tt $, the graph of $w(\delta)$ as shown in Figure (\ref{fig:graph1}) shows a `turning' behaviour at order $\delta\sim  2.06\tt$. For the graph of $\tt\approx 10^{-66}$ the $w(\delta)$ shows a similar turning point.  For $\delta> 2.06 \tt$ the graph is almost a straight line, and the usual classical equation is restored. We could interpret this as a breakdown of the classical approximation.  This is justified in hindsight as the $b_c$ is corrected to order $\tt$ and therefore the angle of scattering stops at a finite value. After this the metric's behaviour for the solution of geodesic changes. We refrain from commenting on the interpretation of the results, but we define a quantum integer 
$n$ such that $\exp(-2 n \pi)= \tt$ or $2 n \pi = -\ln \tt$ at which the photon orbits circulation of the horizon ceases, and the critical radius is reached.
As $\delta >\tt$, the linear behaviour is retained, the straight line has a slightly different slope and an intercept than the classical graph, however as $\delta\sim \tt$, the graph starts deviating. Note, we require a more rigorous calculation than this to identify the quantum behaviour of photons. This is a calculation to show that semiclassical fluctuations are important for photon trajectories with high `n' number of circles around the horizon.

 \begin{figure}
\subfigure{\includegraphics[scale=0.3]{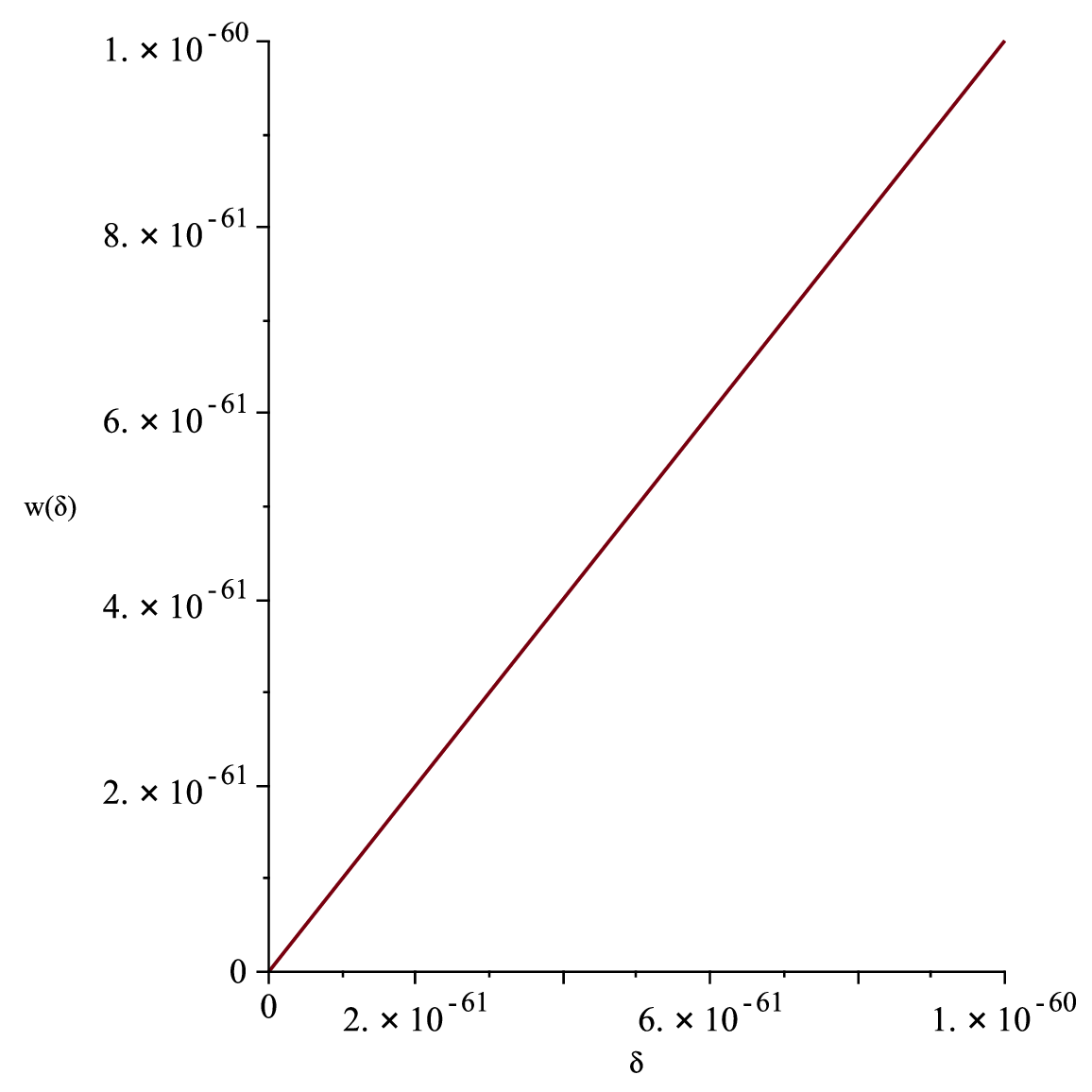}}
\subfigure{\includegraphics[scale=0.3]{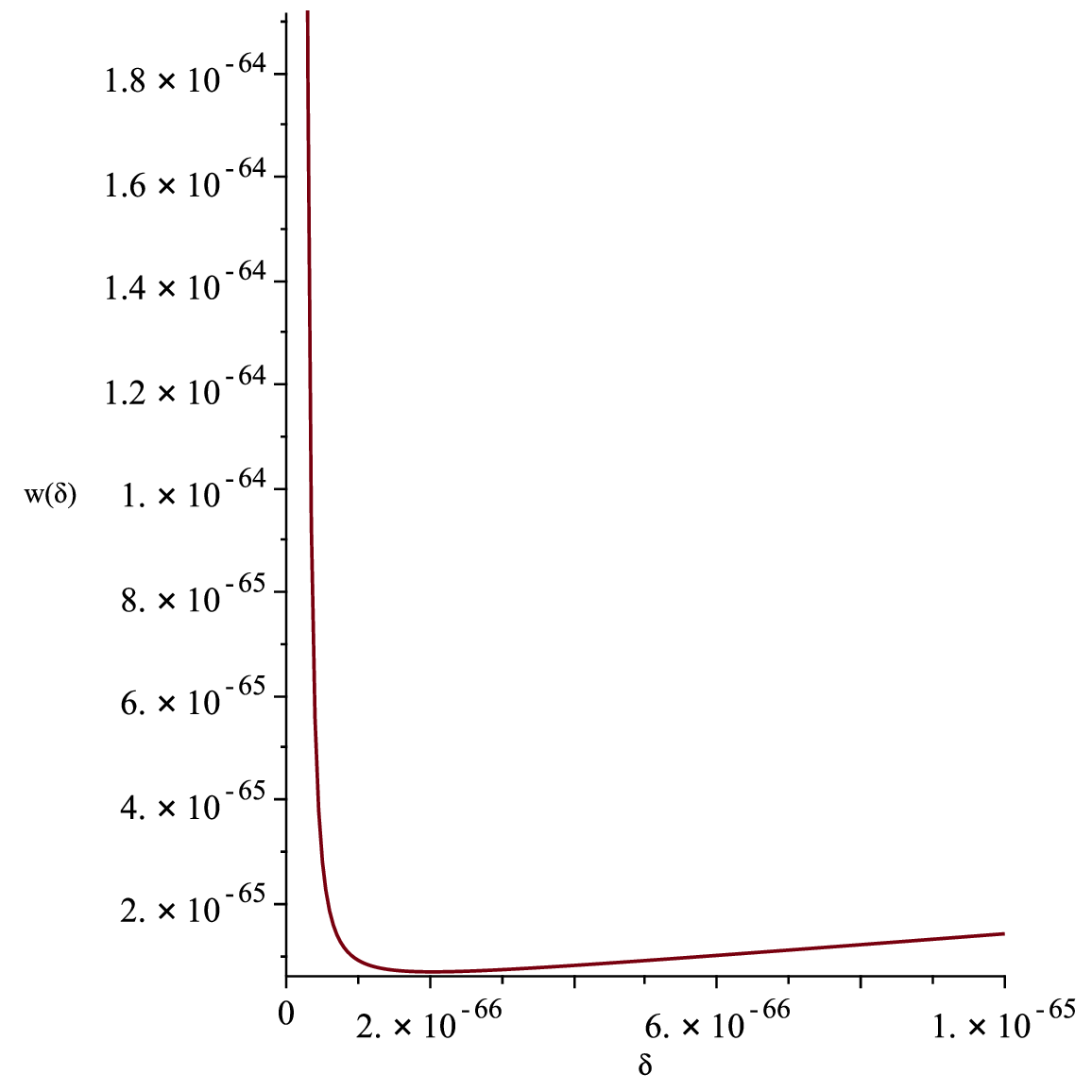}}
 \caption{ $\tt\sim 10^{-66}$ in two different ranges of the y-axis }
 \end{figure}

\section{Conclusion}
Due to the fact that the distance between the photon orbit impact parameters diminish to tiny scale times the black hole mass; we find that the classical behaviour is almost of the order of a semi-classical parameter for coherent states used in \cite{adg}. This motivates us to find if the corrections to the trajectories studied in \cite{adg} will be relevant in the observation of the EHT image, i.e. if there are any further physical implications. We find a finite orbit number `n' for the photons after which the impact parameter deviates again from the critical value, or there is a `bounce'. We show that this might be a sign of quantization $2 n \pi= - \ln (\tt)= \ln(1/\tt)=\ln (10^m)=m \ln(10) $ the exponent of the semiclassical parameter. There is a finite integer $n$ or number of times a photon can encircle a horizon. The physical implications of this on a photographic plate or the image of the horzion are under investigation. As the results are dependent on the semiclassical theory (LQG), the actual image observed might be different, depending on the theory used to find (\ref{eqn:corrf}). We are currently investigating the details of the image formation.

\noindent
{\bf Acknowledgement:} SM was supported by a MITACS summer internship.
\section{Appendix}
In this section we compute the relevant non-zero terms in Equations(\ref{eqn:phi_1}, \ref{eqn:correction}) of $I(u_2)$ and $E_1$ which remain non-zero as $\delta\rightarrow 0$. For the $I(u_2)$, we find 
the integral in Equation(\ref{eqn:integral123}) first. That is labelled as $\tilde I$
\be
\tilde{I}= \frac{2}{b_0^2} \int_0^{u^0_2} \frac{(r_g u - 2 r_g^2 u^2 + 2 r_g^3 u^3 - r_g^5 u^5) ~ du}{(1-r_g u)(u^0_1-u)^{3/2} (u^0_2-u)^{3/2}(u-u^0_3)^{3/2}} 
\label{eqn:integral}
\ee
Using a factorization, the integral becomes
\be
= \frac{2}{b_0^2} \int_0^{u^0_2} \frac{r_g u[1- r_g u + r_g^2 u^2 + r_g^3 u^3]}{(u^0_1-u)^{3/2} (u^0_2 - u)^{3/2} (u-u^0_3)^{3/2}} d u.
\ee

The integral requires Elliptic functions, and we use MAPLE to compute. However, to enable MAPLE to give the answer correctly, we had to redefine the variables.
We use
\be
x^2= k_0 \frac{u^0_2-u}{u^0_1 -u}   \   \    \     \    \    \   k_0=\frac{u^0_1-u^0_3}{u^0_2-u^0_3}
\ee
Using this substitution we get the integral to be of the form
\be
\Lambda r_g \int_0^{\sqrt{k_0 \frac{u^0_2}{u^0_1}}} \frac{ a x^8 + b x^6+ c x^4+ d x^2 + f}{ x^2(1-x^2)^{3/2} (h-x^2)^{3/2}} dx 
\ee
where $\Lambda,a,b,,c,d,f,h$ are functions of $u^0_1,u^0_2,u^0_3$.  We also have to introduce a parameter $e$ as there is a new divergence at $x=0$. We thus take the lower limit as $x=e$ to regulate the divergence.
The result of the integral using MAPLE is 
{\small
\begin{eqnarray*}
&-\frac{1}{(h-1)^2} \left[2\left\{-\frac{(h-1)}{h^{3/2}}\left(a h^3 + \left(-\frac a2 + \frac b2\right)h^2 + \left(\frac{c}2 + \frac{d}{2} + f\right) h - \frac{f}{2} \right) F\left(x,\sqrt{\frac1{h}}\right) \right. \right.&\\
&+ E\left(x,\sqrt{\frac1h}\right) \left(a h^4 + \left(\frac b2 -a\right) h^3 + \left(a + \frac{b}{2} + c + \frac{d}{2} + f\right) h^2 +\left(\frac d2-f\right) h +f\right) \left.\right\}&\\
&+\frac{1}{x\  h^2 \sqrt{-x^2+1}\sqrt{-x^2+h} } \left\{h^4\left(\frac12 a x^4 -\frac12 x^2 a\right) + \left(\frac{bx^4}{2} + \left(-f -\frac a2 -b -\frac{c}2 -\frac d2\right)x^2 + \frac{f}{2}\right)h^3 \right.&\\ &+ \left(\left(\frac a2 + \frac b2 + c + \frac d2+ f\right) x^4 
\left.\left.\left(-\frac{c}2 + \frac{f}{2} \right)x^2 - f\right) h^2+ \frac{(x-2)((d-2f)x^2 -f)(x+1)h}{2} + f x^4 -f x^2 \right\}\right]&
\end{eqnarray*}}

Upon computing the integral and writing the explicit values of $u_1,u_2,u_3$ the answer is obtained using MAPLE. The coefficients are individually
\bea
a&=& \frac{29}{81 M} + \frac{59 }{81 M} \delta - \frac{97}{243 M} \delta^2 \\
b&=& -\frac{116}{81 M} - \frac{818}{243 M}\delta +\frac{1636}{729 M} \delta^2 \\
c&=& \frac{58}{27 M} + \frac{464}{81 M} \delta - \frac{250}{81 M} \delta^2 \\
d&=& -\frac{116}{81 M} - \frac{346}{81 M} \delta -\frac{28}{27 M} \delta^2 \\
f&=& \frac{29}{81 M} + \frac{287}{243 M} \delta + \frac{149}{729 M} \delta^2 \\
h&=& 1 + \frac{4}{3} \delta - \frac{8}{9} \delta^2  \\
\Lambda &=& - 9 \sqrt{2} \frac{M^{7/2}}{\delta^2} \left(1+\frac{4}{3}\delta\right)
\eea



From $|tilde{I}$ we subtract the second integral in Equation (\ref{eqn:tilda}) 
\be
\left(\frac{\tilde{b}}{{ M b_0}^3}-\nu G'(u_2^0)\right)\int_{0}^{u_2^0} \frac{du}{G(u)^{3/2}} 
\ee
This can be transformed using the same variables and Equation(\ref{eqn:b}) as above to a integral of the form
 
 \be
 H(u_2^0) \Lambda \int \frac{(x^2-h)^{5/2}}{x^2 (1-x^2)^{3/2}}  dx
\ee
where 
\be
M b_0^2 (H(u_2^0)) = \left(\frac{58}{81}-\frac{118}{81}\delta^2 + \frac{250}{81}\delta^2\right)
\ee
The result of the integrals is function of Elliptical Integrals as shown by MAPLE.
 \bea
\frac{2}{h }\left( (h-\frac12) (h-1)F\left(x,\sqrt{\frac1h}\right) +h(h^2-h+1) E\left(x,\sqrt{\frac1h}\right)  \right)&& \nn \\
-\frac2{x \sqrt{1-x^2}} \ \sqrt{h-x^2}\left( \left(h^2-h+\frac12\right) x^2-\frac{h^2}{2}\right)&&
\eea

 (The worksheet is available on request). 

For both the integral results when we plugin the limits, in the integral there are also terms of the form $E(e,k)$ and $F(e,k)$ which we use the small $e$ expansion of the Elliptic functions \cite{karp}.

\be
F(x,k)= \sum_{m=0}^{\infty} \frac{x^{2m+1}}{2m+1} _2F_1(-m,1/2;1;1-k^2)
\ee
we keep the $m=0$ term, which gives  for small $x$
\be
F(x,k) \approx x 
\ee

For the upper limit, Elliptic $E(x,k)$, for the $x\approx 1 , k\approx 1$ we use the asymptotic forms as given in \cite{vel}:

\be
E(x,k)= E(k) - \frac{2}{\pi}\left( K'(k')-E' (k')\right) \sinh^{-1} \left(\frac{1}{k'\tan\phi}\right) + (1-k'^2 \sin^2\phi) \cot\phi + O(1+\tan^2\phi)^{1/2} \cot^2\phi(d_0'-..
\ee  
where $x=\sin\phi$.  The constants $d_0'$ are functions of $k'$ and tend to zero as $k'\rightarrow 0$.

where $K(k), E(k)$ is the complete Elliptic functions of the first and second kind, and $F(a,b;c,z)$ is the Hypergeometric function, a series on positive powers of $z$.

For the Elliptic $F(\phi,k)$ integral we have
\be
F(\phi,k)= K-\frac{2}{\pi} K' \sinh^{-1} \left(\frac{1}{k'\tan\phi}\right) + (1+ k'^2 \tan^2\phi)^{1/2} \cot^2\phi(c_0'-..) 
\label{eqn:EllipticK}
\ee
and $c_0'$ is dependent on $k'$ and tends to zero as $k'\rightarrow 0$. The first term in Equation(\ref{eqn:EllipticK}) has singularities proportional to $\ln \left(4/k'\right)$  as $k \rightarrow 1$. 

We find the that the incomplete integrals can be approximated using the above as
\bea
F(\phi,k)&\approx &\ln\frac{4}{k'} -\sinh^{-1}\left(\frac{1}{k'\tan\phi}\right) \label{eqn:ellipf}\\
&& \approx \ln 4 -\frac12 \ln \delta - \ln (1+\delta) -\ln\left(\frac{2}{\sqrt{3}}\right) -\ln\left(\frac1{\sqrt{2}}\left(1+ \sqrt{3}\right) + \frac{\delta}{2\sqrt{2}}\left(1- \frac1{\sqrt{3}}\right)\right)\nn
\eea
and
\be
E(\phi,k) \approx 1+ \frac{2 \delta}{3}(1-\delta)\left(\ln \frac{2\sqrt{3}}{\sqrt{\delta}} -\ln(1-\delta)-\frac12\right)\label{eqn:ellipe}
\ee

The terms from the Integrals which contribute to the equation as $\delta \rightarrow 0$ are the following (MAPLE worksheet is available on request):
\bea
\Gamma(\delta)=\frac{-113280 \delta^2 \ (E(\phi,k)-E(e,k))}{ \sqrt{9+12\delta - 8 \delta^2} \  \eta(\delta)} + \frac{92160 \delta^2 \ (F(\phi,k)-F(e,k))}{\sqrt{9+12\delta - 8\delta^2} \ \eta(\delta)}   && \nonumber \\
-\frac{4864\  \delta^2\ \sqrt{3} \sqrt{24 \delta - 40\delta^2 }}{\sqrt{24 \delta - 24 \delta^2  }\  \eta(\delta)} - \frac{101952 \delta\  (E(\phi,k)-E(e,k))}{\eta(\delta)} - \frac{94720 \sqrt{1-e^2}\ \delta^3}{e\ \sqrt{9+12\delta-9 e^2 -8 \delta^2}\ \eta(\delta)}
&&\label{eqn:expansion}\eea
where $\eta(\delta)= (3\delta-2)(3 \delta^2 - 2\delta + 12)(9+12\delta-8 \delta^2)^2 (6 - 2\delta + 3\delta^2)(e^2-1)$.
And
\be
\sin\phi={1-\frac13\delta +\frac12 \delta^2}
\ee
and
\be
k=\frac1{\sqrt{1+\frac43\delta-\frac89 \delta^2}}
\ee

Note that the computer algorithm keeps all powers of $\delta$ as generated from the integral. 

Using the approximations for the Equation (\ref{eqn:expansion}) we get to quadratic order in $\delta$
\be
\Gamma(\delta)= -0.7228 \ \delta^2 -2.707\  \frac{\delta^3}{e} + 2.6337 \ \delta^2 [F(\phi,k)-F(e,k)] + (0.648 \ \delta^2 - 2.913\  \delta)[ E(\phi,k)- E(e,k)]
\ee
We plot the $\Gamma(\delta),\Gamma(\delta)/\delta,\Gamma(\delta)/\delta^2$ to identify the divergence as $\delta\rightarrow 0$ . From the figures we find that there is a divergence in the integral $I(u)$ as $\delta \rightarrow 0$ which will contribute to the formula for the critical impact parameter, thus we keep this in the semiclassical formula. 

 \begin{figure}
\subfigure{\includegraphics[scale=0.3]{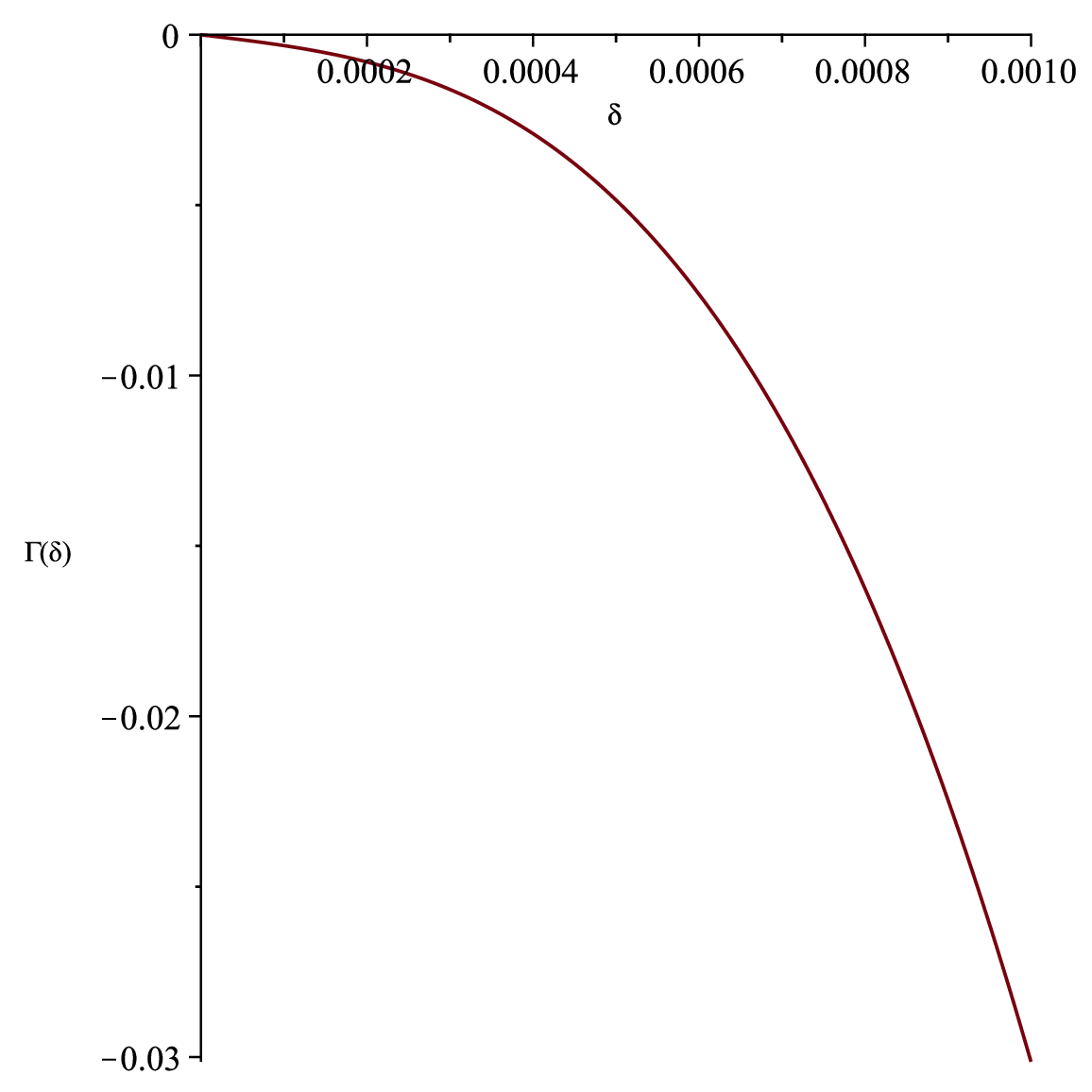}}
\subfigure{\includegraphics[scale=0.3]{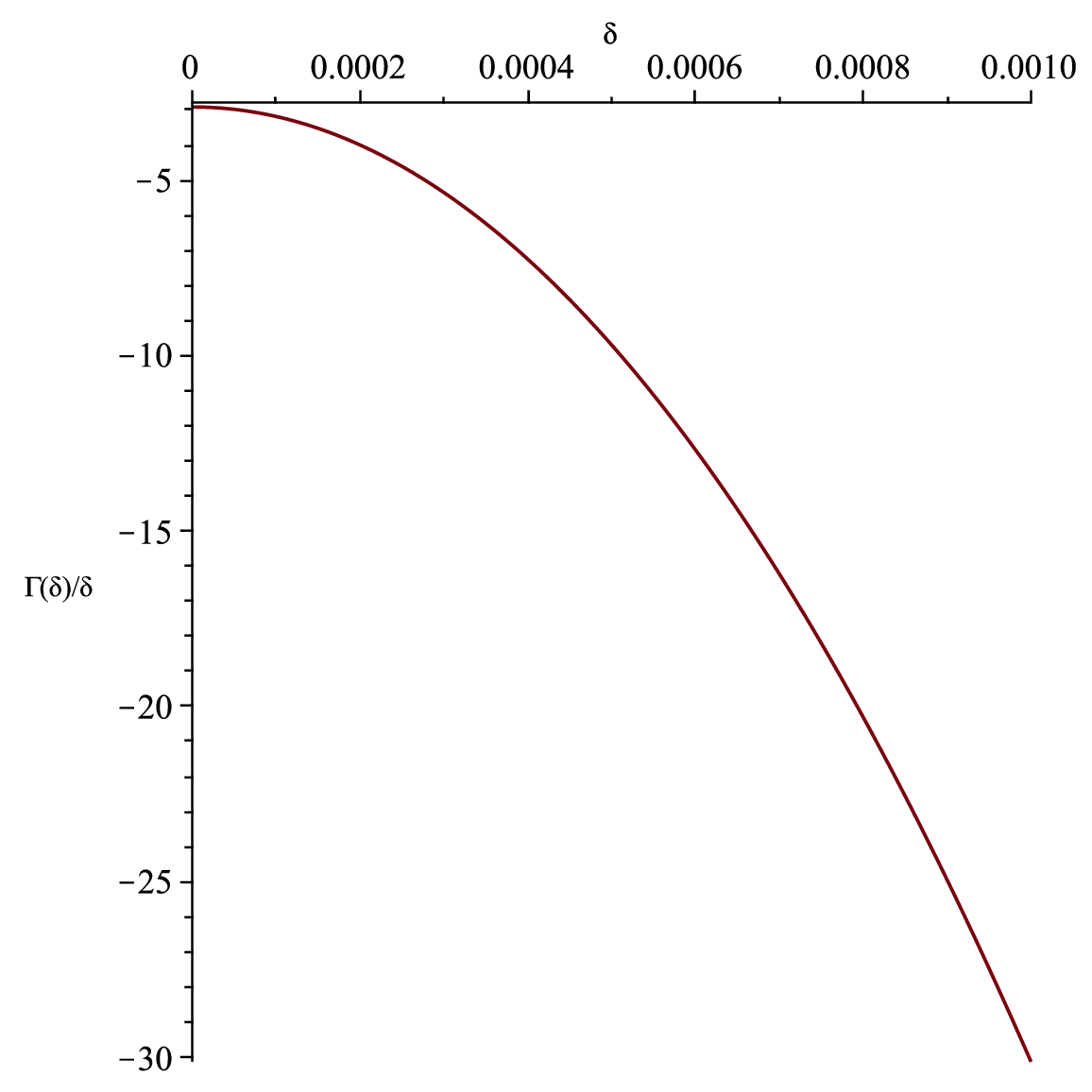}}
\subfigure{\includegraphics[scale=0.3]{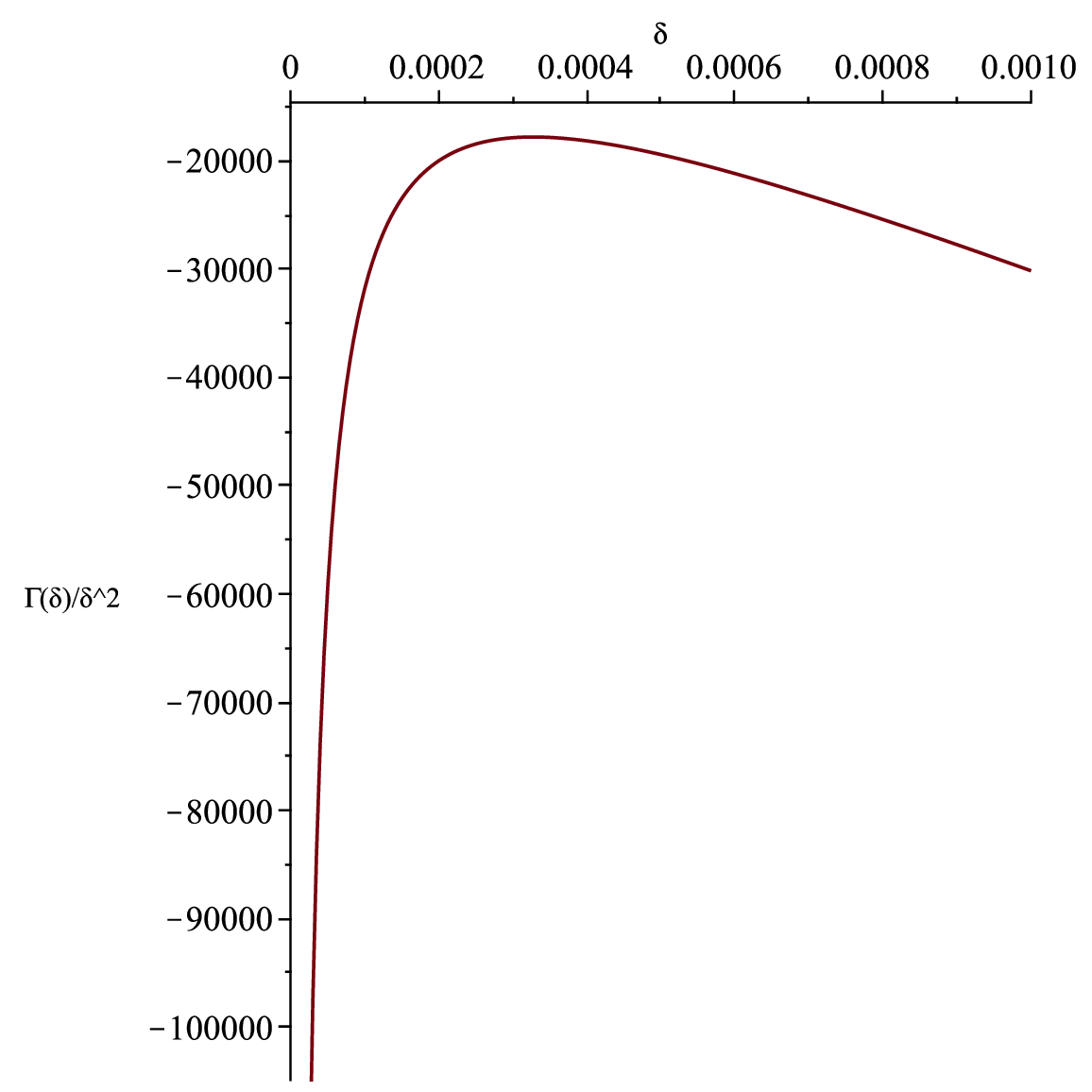}}
 \caption{ Plot of $\Gamma(\delta),\Gamma(\delta)/\delta,\Gamma(\delta)/\delta^2$ }
 \end{figure}

These are then approximated using the Elliptic integrals Equations (\ref{eqn:ellipf}, \ref{eqn:ellipe})


The contribution from $I(u_2)$ is therefore

\bea
&& \frac1{2\sqrt{2M}}  \frac{1}{(2M)^{3/2} b_0^2} \Lambda\left(-0.021 \ \delta^2 - 2.707 \frac{\delta^3}{e} -0.3455 \delta^2 \ \ln \delta -2.913 \delta\right)\\
&=& \frac{1}{12 \sqrt{2}}\left(3.906 + 2.707\  \frac{\delta}{e}  + 0.3455\  \ln\delta + \frac{2.913}{\delta} \right) \eea

Thus in total non zero contribution  when $\delta \rightarrow 0$ is 
\be
\frac{1}{2\sqrt{2M}} I(u_2) \approx 0.23 +0.159 \frac{\delta}{e} +\frac{1.72}{\delta}+0.0203 \ln\delta
\label{eqn:I(u)}
\ee

We neglect the $\delta/e\sim 1$ term in the final result, as that is multiplied by $\tilde{t}$ and gives an infinitesimal contribution comparatively.

For the $E_1$,  defined in Equation(\ref{eqn:phi_1},\ref{eqn:correction}) we simply approximate that as 
\be
\frac{(u_2-u_2^0)}{\sqrt{G_0(u_2)}}= -\frac{\sqrt{u_2^0-u_2}}{\sqrt{(u_1^0-u_2)(u_2-u_3^0)}}\approx -  \sqrt{\frac{0.225 \tt}{0.67 \ \delta + 0.225\  \tt}}\sqrt{2M} \label{eqn:E_1}
\ee
This gives a rather strange $\tt^{1/2}$ dependence, but we keep it for the calculation of the $\phi_{\infty}$ as a function of $\delta$.


\begin{thebibliography}
{99}
\bibitem{eht} The Event Horizon Collaboration, Astrophysical Journal Letters {\bf 875} (2019) 3. 
 \bibitem{adg}A. Dasgupta,  Journal of Cosmology and Astroparticle Physics, {\bf 05} 011 (2010).
 \bibitem{luminet} J. P.  Luminet,  Astronomy and Astrophysics, {\bf 75} 228 (1979). 
 \bibitem{chandra} S. Chandrasekhar, {\it The Mathematical Theory of Black Holes}, Clarendon Press, Oxford University (1991).
 \bibitem{adg1} A. Dasgupta, Journal of Cosmology and Astroparticle Physics {\bf 0308} 004 1 (2003).
\bibitem{adg2} A. Dasgupta, Canadian Journal of Physics {\bf 96} (4)  366 (2018).
\bibitem{adg3} A. Dasgupta, Classical and Quantum Gravity {\bf 23}  635-671 (2006).
 \bibitem{karp} D. Karp et al arXiv:math/0410009
 \bibitem{vel} H. Van de Vel `On the series Expansion Method for computing Incomplete Elliptic Integrals of the first and second kind'  https://www.ams.org/journals/mcom/1969-23-105/S0025-5718-1969-0239732-8/S0025-5718-1969-0239732-8.pdf 
 \end{thebibliography}
\end{document}